\documentclass{naturewithfigure}
\usepackage{graphicx}

\title{Tunable Quantum Beam Splitters for Coherent Manipulation of a Solid-State Tripartite Qubit System}

\author{Guozhu Sun$^{1,2,3\ast}$, Xueda Wen$^{3}$, Bo Mao$^{2}, $ Jian Chen$^{1,3}$, Yang Yu$^{3\ast}$, Peiheng Wu$^{1,3}$, Siyuan Han$^{1,2\ast}$ \\}

\begin{document}

\maketitle

\begin{affiliations}
 \item Research Institute of Superconductor Electronics, School of Electronic Science and Engineering, Nanjing University, Nanjing 210093, China
 \item Department of Physics and Astronomy, University of Kansas, Lawrence, KS 66045, USA
 \item National Laboratory of Solid State Microstructures, School of Physics, Nanjing University, Nanjing 210093, China
\end{affiliations}

\begin{abstract}
Coherent control of quantum states is at the heart of implementing solid-state quantum processors and testing quantum mechanics at the macroscopic level. Despite significant progress made in recent years in controlling single- and bi-partite quantum systems, coherent control of quantum wave function in multipartite systems involving artificial solid-state qubits has been hampered due to the relatively short decoherence time and lacking of precise control methods. Here we report the creation and coherent manipulation of quantum states in a tripartite quantum system, which is formed by a superconducting qubit coupled to two microscopic two-level systems (TLSs). The avoided crossings in the system's energy-level spectrum due to the qubit-TLS interaction act as tunable quantum beam splitters of wave functions. Our result shows that the Landau-Zener-St\"{u}ckelberg interference has great potential in the precise control of the quantum states in the tripartite system.

\end{abstract}

As one of three major forms of superconducting qubits \cite{RevModPhys.73.357,PhysicsToday.58.11,Nature.453.1031}, a flux-biased superconducting phase qubit \cite{QuantumInfProcess.8.81,QuantumInfProcess.8.117} consists of a superconducting loop
with inductance $L$ interrupted by a Josephson junction (Fig. 1a). The
superconducting phase difference $\varphi $ across the junction serves as
the quantum variable of coordinate. When biased close to the critical current $I_{0}$, the
qubit can be thought of as a tunable artificial atom with discrete energy
levels that exist in a potential energy landscape determined by the circuit
design parameters and bias (Fig. 1b). The ground state $|0\rangle $ and the
first excited state $|1\rangle $ are usually chosen as the computational basis states of the phase qubit. The
energy difference between $|1\rangle $ and $|0\rangle ,$ $\omega _{10},$
decreases with flux bias. A TLS is phenomenologically understood to
be an atom or a small group of atoms tunneling between two lattice
configurations inside the Josephson tunnel barrier, with different wave functions $|L\rangle $ and $|R\rangle $ corresponding to different critical current (Fig. 1c). Under the interaction picture of the qubit-TLS
system, the state of the TLS can be expressed
in terms of the eigenenergy basis with $|g\rangle $ (the ground state) and $%
|e\rangle $ (the excited state). When
the energy difference between $|e\rangle$ and $|g\rangle$, $\hbar\omega _{TLS}=E_{e}-E_{g},$ is close
to $\hbar\omega _{10}$ ($\hbar \equiv h/2\pi$ where $h$ is Planck's constant.), coupling between the phase qubit and the TLS becomes significant, which could result in
increased decoherence \cite{QuantumInfProcess.8.81,QuantumInfProcess.8.117}. On the other hand, one can exploit strong qubit-TLS
coupling for demonstrating coherent macroscopic quantum phenomena and/or quantum information processing \cite{PhysRevLett.97.077001,PhysRevLett.100.113601,NaturePhysics.4.523}. For instance, recently a tetrapartite system formed by two qubits, one cavity and one TLS has been studied \cite{QuantumInfProcess.8.117}. However, although multipartite spectral property and vacuum Rabi oscillation have been observed coherent manipulation of the quantum states of the whole system has not been demonstrated yet.

In our experiments, we use two TLSs near 16.5 GHz to form a hybrid tripartite \cite{PhysRevLett.94.027003,PhysRevA.62.062314,ChristianF.Roos06042004} phase qubit-TLS system and demonstrate Landau-Zener-St\"{u}ckelberg (LZS) interference in such tripartite system. The avoided crossings due to the qubit-TLS interaction act as tunable quantum beam splitters of wave functions, with which we could precisely control the quantum states of the system.

\noindent\textbf{Results}

\noindent\textbf{Experimental results of LZS interference.} Fig. 1d shows the measured spectroscopy of a phase qubit. The spectroscopy data clearly show two avoided crossings
resulting from qubit-TLS coupling. Since after the application of the $\pi$-pulse the
system has absorbed exactly one microwave photon and the subsequent steps of state manipulation are
accomplished in the absence of the microwave, conservation of energy guarantees that one and only
one of the qubit, TLS1 and TLS2, can be coherently transferred to its excited state. Thus only  $%
\{|1g_{1}g_{2}\rangle ,|0e_{1}g_{2}\rangle ,|0g_{1}e_{2}\rangle \}$ as marked in Fig. 1d are involved in the dynamics of the system. Notice that these three basis states form a generalized W state \cite{PhysRevA.62.062314,ChristianF. Roos06042004,Nature.438.643}, $|\psi \rangle =\alpha |1g_{1}g_{2}\rangle +\beta
|0e_{1}g_{2}\rangle +\gamma |0g_{1}e_{2}\rangle ,$ which preserves entanglement between the remaining bipartite system even when one of the qubit is lost and has been recognized as an important resource in quantum information science \cite{Nature.404.247}.
The system's effective Hamiltonian can be
written as%
\begin{equation}
H=\hbar \left(
\begin{array}{cccc}
\omega _{10}(t) & \Delta _{1} & \Delta_{2} &  \\
\Delta _{1} & \omega _{TLS1} & 0 &  \\
\Delta _{2} & 0 & \omega _{TLS2} &
\end{array}%
\right) , \label{Hamiltonian1}
\end{equation}%
where $\Delta _{1}$ ($\Delta _{2}$) is the coupling strength
between the qubit and TLS1 (TLS2). $\omega _{TLS1}$ ($\omega
_{TLS2}$) is the resonant frequency of TLS1 (TLS2). $\omega _{10}(t)=\omega
_{10,dc}-s\Phi(t)$, with $\omega _{10,dc}$ being the initial energy detuning
controlled by the dc flux bias line (i.e., the second platform holds in the dc
flux bias line), $s=|d\omega _{10}(\Phi)/d\Phi|$ being the diabatic energy-level
slope of state $|1g_{1}g_{2}\rangle $, and $\Phi(t)$ being the time dependent
flux bias (Fig. 1a).

In our experiment, coherent quantum control of multiple qubits is realized with
Landau-Zener (LZ) transition. When the system is swept through the
avoided crossing, the asymptotic probability of transmission is $\exp\left(-2\pi \frac{\Delta^{2}}{\nu }\right),$
where $\hbar \nu \equiv dE/dt$ denotes the rate of the energy
spacing change for noninteracting levels, and $2\hbar\Delta $ is the minimum energy gap.
It ranges from 0 to 1, depending on the ratio of $\Delta $ and $%
\nu$. The avoided crossing serves as a
beam splitter that splits the initial state into a coherent superposition of
two states \cite{J.R. Petta02052010}. These two states evolve independently in time while a relative
phase is accumulated causing interference after sweeping back and forth through the avoided crossing.
Such LZS interference has been observed recently in
superconducting qubits \cite%
{WilliamD.Oliver12092005,PhysRevLett.96.187002,PhysRevLett.98.257003,PhysRevLett.101.017003,PhysRevLett.101.190502,sun:102502,Nature.459.960,Shevchenko2010}%
. However, in these experiments the avoided crossings of the single qubit energy spectrum are
used, and microwaves, whose phase is difficult to control, are applied to drive the
system through the avoided crossing consecutively to manipulate the qubit state. Here we use a triangular bias waveform with width shorter than the qubit's decoherence time to coherently control the
quantum state of the tripartite system. The use of a triangular waveform, with a time resolution of 0.1 ns, ensures precise control of the flux bias sweep at a constant rate and thus the quantum state.  The qubit is initially prepared
in $|0g_{1}g_{2}\rangle .$ A resonant microwave $\pi$-pulse is applied to coherently transfer the
qubit to $|1g_{1}g_{2}\rangle $. A triangular flux bias $\Phi(t)$ with variable
width $T$ and amplitude $\Phi_{LZS}$
\begin{equation}
\Phi(t)=\left\{
\begin{array}{rl}
\frac{2\Phi_{LZS}}{T}t, & t<T/2 \\
2\Phi_{LZS}-\frac{2\Phi_{LZS}}{T}t, & T/2<t<T,%
\end{array}%
\right.   \label{pulse}
\end{equation}%
is then applied immediately to the phase qubit to induce LZ transitions (Fig. 2d). This is followed by a short
readout pulse (about 5 ns) to determine the probability of finding the qubit in the state $|1\rangle $, i.e., the system in the state $|1g_{1}g_{2}\rangle $.

Fig. 2a shows the measured population of $|1\rangle $ as a function
of $T$ and $\Phi_{LZS}$. On the top part of the plot,
the amplitude is so small that the state could not reach the first avoided
crossing $M_{1}$. Therefore, no LZ transition could occur and only trivial
monotonic behavior is observed. When the amplitude is large enough to reach %
$M_{1}$, the emerging interference pattern can be qualitatively divided into
three regions with remarkably different fringe patterns.

\noindent\textbf{Quantitative comparison with the model.} To quantitatively model the data, we calculate the probability to return to the initial
state $P_1$ by considering the action of the unitary operations on the initially prepared state.
Neglecting relaxation and dephasing, we find
\begin{equation}
\begin{array}{ll}
P_1=(1-P_{LZ1})^2+P_{LZ1}^2(1-P_{LZ2})^2+P_{LZ1}^2P_{LZ2}^2\\
-2P_{LZ1}(1-P_{LZ1})(1-P_{LZ2})\cos(\theta_I+2\tilde{\theta}_{S1}-2\tilde{\theta}_{S2})\\
-2P_{LZ1}^2P_{LZ2}(1-P_{LZ2})\cos(\theta_{II}+2\tilde{\theta}_{S2})\\
+2P_{LZ1}(1-P_{LZ1})P_{LZ2}\cos(\theta_I+\theta_{II}+2\tilde{\theta}_{S1}),
\end{array}%
\label{P1}
\end{equation}%
where $P_{LZi}$ ($i=1,2$) is the Landau-Zener transition probability at the $i$th avoided crossing $M_i$, and $\theta_I$ and $\theta_{II}$ are the phases accumulated in region $I$ and $II$, respectively (Fig.2b).
The phase jump $\tilde{\theta}_{Si}=\theta_{Si}-\pi/2$ ($i=1,2$) at the $i$th avoided crossing
is due to the Stokes phase \cite{   PhysRevLett.96.187002,Shevchenko2010} $\theta_{Si}$ which depends on the adiabaticity parameter $\eta_i=\Delta_i^2/\nu$ in the form
$\theta_{Si}=\pi/4+\eta_i(\ln\eta_i-1)+\arg\Gamma(1-i\eta_i)$, where $\Gamma$
is the Gamma function. In the adiabatic limit $\theta_S\to0$, while in the sudden limit $\theta_S=\pi/4$. In order to give a clear
physical picture, hereafter we adopt the terminology of optics to discuss the phenomenon
and its mechanism. First of all we define two characteristic sweeping rates of $\nu _{1}
$ and $\nu _{2}$ from $2\pi \Delta _{i}^{2}/\nu _{i}= 1$ ($i=1,2$). From the spectroscopy data, we have $\Delta
_{1}/2\pi=10$ MHz and $\Delta _{2}/2\pi=32$ MHz, thus $\nu _{1}/2\pi= 3.94\times
10^{-3}$ GHz/ns and $\nu _{2}/2\pi= 4.04\times 10^{-2}$ GHz/ns, respectively.
These lines of constant sweeping rate characteristic to the system are marked as oblique dotted lines in Fig. 2a.
The avoided crossings $M_{1}$ ($M_{2}$) can be viewed
as wave function splitters with controllable transmission coefficients set by
the sweeping rate $\nu $. $\nu _{1}$ and $\nu _{2}$ thereby define three regions in the $T-\Phi_{LZS}$ parameter plane that contain all main features of the measured
interference patterns:

(I) $\nu \simeq \nu _{1}$ and $\nu \ll \nu _{2}$: $M_{1}$ acts as a beam
splitter and $M_{2}$ acts as a total reflection mirror, i.e., $P_{LZ1}\simeq 1/2$ and $P_{LZ2}\simeq 0$.
In this case, equation (\ref{P1}) can be simplified as
\begin{equation}
P_1=1-2P_{LZ1}(1-P_{LZ1})[1+\cos(\theta_{I}+2\tilde{\theta}_{S1}-2\tilde{\theta}_{S2})].
\end{equation}
Apparently, only path $\#1$ and path $\#2$ contribute to the interference. The phase
accumulated in region $I$ can be expressed as
\begin{equation}
\theta_I=\int_{0}^{T}[\omega _{1}(t)-\omega _{2}(t)]dt,
\label{Phase1}
\end{equation}%
where $\omega_{i}(t)$ ($i=1,2$) denotes the energy frequency corresponding to path $\#i (i=1,2)$. It is easy to find that $P_1$ is maximized (constructive interference) in the condition
\begin{equation}
\theta _{total}=\theta_I+2\tilde{\theta}_{S1}-2\tilde{\theta}_{S2}=(2n+1)\pi,\  (n=0,1,2,\cdots ),  \label{Phase2}
\end{equation}
from which we can obtain the analytical expression for the positions of
constructive interference fringes
\begin{equation}
\left\{
\begin{array}{ll}
\frac{1}{2}s\Phi_{LZS}\left(1-\frac{\delta _{1}}{s\Phi_{LZS}}\right)^{2}T+2\tilde{\theta}_{S1} =(2n+1)\pi , & \delta
_{1}<s\Phi_{LZS}<\delta _{2} \\
\frac{1}{2}\left[\frac{\delta _{12}^{2}}{s\Phi_{LZS}}+2\left(1-\frac{\delta _{2}}{s\Phi_{LZS}}\right)\delta
_{12}\right]T+2(\tilde{\theta}_{S1}-\tilde{\theta}_{S2})=(2n+1)\pi , & s\Phi_{LZS}>\delta _{2}%
\end{array}%
\right.
\end{equation}%
\\
where $\delta _{1}=\omega _{10,dc}-\omega _{TLS1}$, $\delta _{2}=\omega
_{10,dc}-\omega _{TLS2}$, and $\delta _{12}=\omega _{TLS1}-\omega _{TLS2}$.

In Fig. 2b we show the calculated constructive interference strips which
agree well with the experimental result. Especially, in the limit of $s\Phi_{LZS}\gg
\delta _{2}, \delta _{12}$, equation (7) can be simplified as
\begin{equation}
\delta _{12}T+2(\tilde{\theta}_{S1}-\tilde{\theta}_{S2})=(2n+1)\pi.
\end{equation}
Intuitively, this result is straightforward to understand since in the
large amplitude limit the accumulated phase $\theta _I$ is two times
the area of a rectangle with length $T/2$ and width $\omega _{TLS1}-\omega
_{TLS2}$.

(II) $\nu \simeq \nu _{2}$ and $\nu \gg \nu _{1}$: $M_{1}$ acts as a total
transmission mirror and $M_{2}$ acts as a beam splitter,
i.e., $P_{LZ1}\simeq1$ and $P_{LZ2}\simeq 1/2.$
In this case, equation (\ref{P1}) can be simplified as
\begin{equation}
P_1=1-2P_{LZ2}(1-P_{LZ2})[1+\cos(\theta_{II}+2\tilde{\theta}_{S2})].
\end{equation}
Only path $\#2$ and path $\#3$ contribute to the interference. Using the same
method in dealing with the region I, we obtain the analytical formula governing
the positions of constructive interference fringes:
\begin{equation}
\frac{1}{2}s\Phi_{LZS}\left(1-\frac{\delta _{2}}{s\Phi_{LZS}}\right)^{2}T+2\tilde{\theta}_{S2}=(2n+1)\pi.  \label{position}
\end{equation}
As shown in Fig. 2c, the
positions of the constructive interference fringes obtained from equation (\ref%
{position}) agree with experimental results very well. Similarly, in
the limit $s\Phi_{LZS}\gg \delta _{2}$, equation (\ref{position}) has the simple form
\begin{equation}
\frac{1}{2}s\Phi_{LZS}T+2\tilde{\theta}_{S2}=(2n+1)\pi ,
\end{equation}
which is also readily understood because in the large amplitude
limit the accumulated phase $\theta _{II}$ is two times the area of a
triangle with base-length $T/2$ and height $s\Phi_{LZS}$.

(III) $\nu _{1}< \nu < \nu _{2}$. This region is more interesting and complex.
Here, $M_{1}$ acts as a beam splitter while $M_{2}$
can act either as a beam splitter or a total reflection mirror. This effect
cannot be described by the asymptotic Landau-Zener formula
because in this region LZS interference occurs only in a relatively small range around
the avoided crossings. Since the analytical solution is extremely complicated
which does not provide clear intuition about the underlying physics, we
use a numerically calculated LZ transition probability $P_{LZ}$ corresponding to the transmission coefficient of $M_{1}$ and $M_{2}$ for comparison with the experimental data.
 We find that for
certain sweeping rates, LZ transition probability resulting from $M_{2}$ is quite low. Therefore, $M_{2}$ can be treated as a total reflection
mirror while $M_{1}$ is still acting as a good
beam splitter. The interference fringes generated by $M_{2}$ thus disappear (the
fringes tend to fade out) and the interference fringes generated by $M_{1}$
dominate, displaying as a chain of
`hot spots' marked by the circles in Fig. 2a.

When both $M_{1}$ and $M_{2}$ can be treated as beam splitters, all three paths (\#1, \#2, and \#3) contribute to the interference. According to equation (\ref{P1}), $P_1$ is maximized in the condition
\begin{equation}
\left\{
\begin{array}{ll}
\theta_I+2(\tilde{\theta}_{S1}-\tilde{\theta}_{S2})=(2n_1+1)\pi, \ \ (n_1=0,1,2\cdots)\\
\theta_{II}+2\tilde{\theta}_{S2}=(2n_2+1)\pi, \ \ (n_2=0,1,2\cdots)
\end{array}.%
\right.
\end{equation}%
It is noted that under this condition the term ($\theta_I+\theta_{II}+2\tilde{\theta}_{S1}$) in equation (\ref{P1})
equals $2n\pi$.
Considering different weights in each path, it is more convenient to obtain a theoretical prediction from a numerical simulation. Here we utilize the Bloch equation to describe the time
evolution of the density operator of the tripartite system:
\begin{equation}
\dot{\rho}=-\frac{i}{\hbar}[\hat{H},\rho ]-\Gamma \lbrack \rho ], \label{Master}
\end{equation}%
where $\Gamma \lbrack \rho ]$ includes the effects of energy relaxation. Fig. 3a shows the calculated population of $|1\rangle $ as a function
of $T$ and $\Phi_{LZS}$. Fig. 3b is the extracted data for different $T$ and $\Phi_{LZS}$. The agreement between
the theoretical and experimental results is remarkable. In order to better
understand the origin of the `hot spots', we also plot the probabilities of LZ transition as a function of
the pulse width at fixed amplitude $\Phi_{LZS}=10 $ m$\Phi _{0}$ (Fig. 3c)$.$
Notice that both LZ transition probabilities oscillate with $T$, which are quite different from the general asymptotic LZ transition probabilities. The transition probability
at $M_{1}$ is always greater because $\Delta _{1}$ is much smaller
than $\Delta _{2}.$ The three oblique dotted lines in Fig. 3a represent lines of constant sweeping rate.
The `hot spots' are located on
these lines, where the transition probability of $M_{2}$ is a minimum. $M_{2}$
thereby acts as a total reflection mirror resulting in the `hot spots' in transition probability.
This feature
further confirms that the avoided crossings play the role of quantum mechanical wave function splitters, analogous to continuously tunable beam splitters in optical experiments. The transmission coefficient of the wave function splitters (the avoided crossings) in our experiment can be varied \emph{in situ} from zero (total reflection) to unity (total transmission) or any value in between by adjusting the duration and amplitude of the single triangular bias waveform used to sweep through the avoided crossings.

\noindent\textbf{Precise control of the quantum states in the tripartite system.} We emphasize that the method of using LZS interference for precise quantum state manipulation described above is performed within the decoherence time of the tripartite system which is about 140 ns. Through coherent LZ transition we can thus achieve a high degree of control over the
quantum state of the qubit-TLS tripartite system. For example, one may take advantage of LZS to control the generalized W state, $|\psi \rangle =\alpha |1g_{1}g_{2}\rangle +\beta
|0e_{1}g_{2}\rangle +\gamma |0g_{1}e_{2}\rangle ,$ evolving in the sub-space spanned by the three product states during the operation of sweeping flux
bias.  In order to quantify the generalized W state, we define
$ w = 1-\sqrt{\sum\limits_{\sigma}(|\sigma|-1/\sqrt{3})^{2}} $, where $\sigma=\alpha,\beta,\gamma$. In Fig. 3d, $w$ is plotted
as a function of $T$ and $\Phi_{LZS}$. Note that with precise control of the flux bias sweep, the states with $w = 1$, which are generalized W states with equal probability in each of the three basis product states, are obtained demonstrating the effectiveness of this new method. It should be pointed out that when one of the three qubits is lost, the remaining two qubits are maximally entangled.

In summary, our tripartite system includes a macroscopic object, which is relatively easy to control and readout, coupled to microscopic degrees of freedom that are less prone to environment induced decoherence and thus can be used as a hybrid qubit. The excellent agreement between our data and theory over the entire $T - \Phi_{LZS}$ parameter plane indicates strongly that the states created are consistent with the generalized W states. The coherent generation and manipulation of generalized W states reported here demonstrates an effective new technique for the precise control of multipartite quantum states in solid-state qubits and/or hybrid qubits\cite{PhysRevLett.97.077001,NaturePhysics.4.523}.

\noindent\textbf{Methods}

\noindent\textbf{Experiment detail.} Fig. 1a shows the principal circuitry of the measurement. The flux bias and microwave are fed through the on-chip thin film flux lines coupled inductively to the qubit. The slowly varying flux bias is used to prepare the initial state of the qubit and to readout the qubit state after coherent state manipulation. In the first platform of the flux bias, the potential is tilted quite asymmetrically to ensure that the qubit is initialized in the left well. Then we increase the flux bias to the second platform until there are only a few energy levels including the computational basis states $|0\rangle $ and $|1\rangle $ in the left well. A microwave $\pi$-pulse is applied to rotate the qubit from $|0\rangle $ to $|1\rangle $. This is followed by a triangular waveform with adjustable width and amplitude applied to the fast flux bias line, which results in LZ transition. A short readout pulse of flux bias is then used to adiabatically reduce the well's depth so that the qubit will tunnel to the right well if it was in $|1\rangle $ or remain in the left well if it was in $|0\rangle $. The flux bias is then lowered to the third platform, where the double well potential is symmetric, to freeze the final state in one of the wells. The state in the left or right well corresponds to clockwise or counterclockwise current in the loop, which can be distinguished by the dc-SQUID magnetometer inductively coupled to the qubit. By mapping the states $|0\rangle $ and $|1\rangle $ into the left and right wells respectively, the probability of finding the qubit in state $|1\rangle $ is obtained. We obtained $T_1\simeq 70 $ ns$ $ from energy relaxation measurement (Fig. S1a), $T_R\simeq 80$ ns$ $ from Rabi oscillation (Fig. S1b), $T_2^*\simeq 60 $ ns$ $ from Ramsey interference fringe (Fig. S1c and Fig. S1d) and $T_2\simeq 137 $ ns$ $ from Spin-echo (Fig. S1e) in the region free of qubit-TLS coupling.

\noindent\textbf{Hamiltonian in our tripartite system.} For the coupled qubit-TLS system, the Hamiltonian can be written as \cite{PhysRevB.72.024526,lupascu:172506}
\begin{equation}
H_{q-TLSs}=H_q+\sum_{i=1}^{2}{H_{TLSi}}+\sum_{i=1}^{2}{H_{q-TLSi}}. \label{Hamiltonian14}
\end{equation}%
In the two-level approximation the effective Hamiltonian of the qubit is $H_q=-\frac{\hbar}{2}\omega_{10}\sigma_z^q$, where the flux bias ($\Phi$) dependent energy level spacing of the qubit $\hbar\omega_{10}=E_1-E_0$ can be obtained numerically by solving the eigenvalues problem associated with the full Hamiltonian of the phase qubit \cite{PhysRevLett.76.3404}. The Hamiltonian of the $i$th TLS can be written as
$H_{TLSi}=-\frac{\hbar}{2}\omega_{TLSi}\sigma_{z}^{TLSi}$, where $\hbar\omega_{TLSi}$ is the energy level spacing of the $i$th TLS. The interaction Hamiltonian between the qubit and the $i$th TLS is
$H_{q-TLSi}=\hbar\Delta_i\sigma_{x}^{q}\otimes\sigma_{x}^{TLSi}$, where $\Delta_i$  is the coupling strength between the qubit and the $i$th TLS and  $\sigma_{x,y,z}^q$ ($\sigma_{x,y,z}^{TLSi}$ ) are the Pauli operators acting on the states of the qubit (the $i$th TLS). By adjusting the flux bias, the qubit and TLSs can be tuned into and out of resonance, effectively turning on and off the couplings. Below  $|0 \rangle$ and $|1 \rangle$  ($|g_i \rangle$  and $|e_i \rangle$ ) are used to denote the ground state and excited state of the qubit (the $i$th TLS). In our experiment the initial state is prepared in the system's ground state $|0g_1g_2 \rangle$. When the couplings between the qubit and TLSs are off, we use a  $\pi$-pulse to pump the qubit to  $|1 \rangle$ (thus the system to $|1g_1g_2 \rangle$. We then sweep the flux bias through the avoided crossing(s) to turn on the coupling(s) between the qubit and the TLS(s).
Since after the application of the $\pi$-pulse the system has absorbed exactly one microwave photon and
the subsequent steps of state manipulation are accomplished in the absence of the microwave, conservation of energy guarantees that one and only one of the qubit, TLS1 and TLS2, can be coherently transferred to its excited state. Therefore, states with only one of the three subsystems in excited state,
$|1g_1g_2 \rangle$, $|0e_1g_2 \rangle$, and $|0g_1e_2 \rangle$, are relevant in discussing the subsequent coherent dynamics of the system. In the subspace spanned by these three basis states the Hamiltonian (\ref{Hamiltonian14}) can be written explicitly as Hamiltonian (\ref{Hamiltonian1}) in the main text.

\noindent\textbf{Unitary operation in our tripartite system.} We use transfer matrix method \cite{PhysRevLett.96.187002,Shevchenko2010} to obtain the probability of finding the system in $|1g_1g_2 \rangle$ at the end of triangular pulse. We use  $|a \rangle$=$[1,0,0]^T$, $|b \rangle$=$[0,1,0]^T$ and $|c \rangle$=$[0,0,1]^T$ to denote the instantaneous eigenstates of the time-dependent Hamiltonian (\ref{Hamiltonian14}), as shown in Fig. S2. It is noted that at the initial flux bias point, which is far from the avoided crossings, the system is in $|a \rangle$=$|1g_1g_2 \rangle$. At the crossing times $t = t_1$ and $t = t_2$, the incoming and outgoing states are connected by the transfer matrix:
\begin{equation}
\hat{U}_1=\left(
\begin{array}{ccc}
\cos(\theta_1/2)\exp(-i\tilde{\theta}_{S1}) & i\sin(\theta_1/2) & 0 \\
i\sin(\theta_1/2) & \cos(\theta_1/2)\exp(i\tilde{\theta}_{S1})  & 0 \\
0 & 0 & 1
\end{array}%
\right)
\end{equation}
and
\begin{equation}
\hat{U}_2=\left(
\begin{array}{ccc}
1 & 0 & 0 \\
0 & \cos(\theta_2/2)\exp(-i\tilde{\theta}_{S2}) & i\sin(\theta_2/2) \\
0 & i\sin(\theta_2/2) & \cos(\theta_2/2)\exp(i\tilde{\theta}_{S2}) \\
\end{array}%
\right)
\end{equation}
respectively. Here $\sin^2(\theta_i/2)=P_{LZi} (i = 1, 2)$ is the Landau-Zener transition probability at the $i$th avoided crossing. $\tilde{\theta}_{Si}=\theta_{Si}-\pi/2$, where $\theta_{Si}$ is the Stokes phase \cite{PhysRevLett.96.187002,Shevchenko2010} whose value depends on the adiabaticity parameter $\eta_i=\Delta_i^2/\upsilon$ in the form of
$\theta_{Si}=\pi/4+\eta_i(\ln{\eta_i-1})+\arg\Gamma(1-i\eta_i)$,
where $\Gamma$ is the Gamma function. In the adiabatic limit $\theta_S\rightarrow 0$, and in the sudden limit $\theta_S=\pi/4$. At crossing times $t = t_3$ and $t = t_4$, we have $\hat{U}_3=\hat{U}_2$ and $\hat{U}_4=\hat{U}_1$, respectively. The outgoing state at $t = t_i$ and the incoming state at $t = t_{i+1} (i = 0, 1, 2, 3, 4)$ is thus connected by the propagator
\begin{equation}
\hat{U}_{i+1,i}=\left(
\begin{array}{ccc}
\exp\left(-i\int_{t_i}^{t_{i+1}}\omega _{a}(t)dt\right) & 0 & 0 \\
0 & \exp\left(-i\int_{t_i}^{t_{i+1}}\omega _{b}(t)dt\right) & 0 \\
0 & 0 & \exp\left(-i\int_{t_i}^{t_{i+1}}\omega _{c}(t)dt\right) \\
\end{array}%
\right)
\end{equation}
where $\omega_i(t)$ is the energy level spacing frequency of  $|i \rangle (i=a,b,c)$ at time $t$. The net effect of a triangular pulse is to cause the state vector to evolve according to the unitary transformation
\begin{equation}
\hat{U}=\hat{U}_{54}\hat{U}_{4}\hat{U}_{43}\hat{U}_{3}\hat{U}_{32}\hat{U}_{2}\hat{U}_{21}\hat{U}_{1}\hat{U}_{10}
\end{equation}%
The probability of finding the system remaining at the initial state is $P_1=|\langle1g_1g_2|\hat{U}|1g_1g_2\rangle|^2$. Its concrete form is equation (\ref{P1}),
where $\theta_I=\int_{t_1}^{t_4}[\omega _{a}(t)-\omega _{b}(t)]dt$  and $\theta_{II}=\int_{t_2}^{t_3}[\omega _{b}(t)-\omega _{c}(t)]dt$ are the relative phases accumulated in region I and II as shown in Fig. S2, respectively. The LZS in our experiment can be viewed as interferences among the three paths, which are labeled with $\#1$, $\#2$, and $\#3$, starting from the same initial state:\\
path $\#1$: $|a\rangle(t<t_1)\rightarrow|a\rangle(t_1<t<t_4)\rightarrow|a\rangle(t_4<t<T)$\\
path $\#2$: $|a\rangle(t<t_1)\rightarrow|b\rangle(t_1<t<t_4)\rightarrow|a\rangle(t_4<t<T)$\\
path $\#3$: $|a\rangle(t<t_1)\rightarrow|b\rangle(t_1<t<t_2)\rightarrow|c\rangle(t_2<t<t_3)\rightarrow|b\rangle(t_3<t<t_4)\rightarrow|a\rangle(t_4<t<T)$\\
Denoting $\omega_i(t)$  as the energy level spacing frequency corresponding to path $i (i = 1, 2, 3)$, then $\theta_I$ and $\theta_{II}$ have the forms  $\theta_I=\int_{0}^{T}[\omega _{1}(t)-\omega _{2}(t)]dt$  and $\theta_{II}=\int_{0}^{T}[\omega _{2}(t)-\omega _{3}(t)]dt$, respectively.

\noindent\textbf{Numerical simulation of LZS interference in bipartite qubit-TLS system.} For the bipartite qubit-TLS system discussed here, the qubit is coupled only to a single TLS. The quantum dynamics of the system, including the effects of dissipation, is described by the Bloch equation of the time evolution of the density operator:
\begin{equation}
\dot\rho=-\frac{i}{\hbar}[\hat{H}_b,\rho]-\Gamma[\rho],\label{blocheq}
\end{equation}
where $\hat{H}_b=\hbar
\left(\begin{array}{cccc}
\omega_{10}(t)              &\Delta   \\
 \Delta                & \omega_{TLS} \\
\end{array} \right)$
with $\omega_{10}(t)=\omega_{10,dc}-\nu t$, $\nu\equiv2s\Phi_{LZS}/T$ is  the energy sweeping rate and $\Delta$ is the qubit-TLS coupling strength. The second term $\Gamma[\rho]$ describes the relaxation process to the ground state $|0g \rangle$ and dephasing process phenomenologically. In a concrete expression, equation (\ref{blocheq}) can be written as: (for ease of discussion, we relabel $|1g \rangle$ and $|0e \rangle$ as $|a \rangle$ and $|b \rangle$, respectively)
\begin{equation}\label{region1b}
\left\{
\begin{array}{rl}
&\dot{\rho}_{aa}=-i\Delta(\rho_{ba}-\rho_{ab})-\Gamma_a\rho_{aa},\\
&\dot{\rho}_{bb}=i\Delta(\rho_{ba}-\rho_{ab})-\Gamma_b\rho_{bb},\\
&\dot{\rho}_{ab}=-i\Delta(\rho_{bb}-\rho_{aa})-i(\omega_{10}(t)-\omega_{TLS})\rho_{ab} \\
&\phantom{\dot{\rho}_{ab}=}-\gamma_{ab}\rho_{ab},\nonumber\\
\end{array}
\right.
\end{equation}
with $\rho_{ba}=\rho^{*}_{ab}$. Here $\Gamma_{\alpha}(\alpha=a,b)$ is the relaxation rate from state $|\alpha\rangle$
to the ground state $|0g\rangle$. The decoherence rate $\gamma_{ab}=(\Gamma_a+\Gamma_b)/2+\gamma^{(deph)}$
includes contributions from both relaxation and dephasing. Fig. S3a and Fig. S3b give the numerically simulated LZS interference pattern for the qubit coupled with the first TLS and second TLS, respectively. To calculate the transmission coefficient of $M_i (i = 1, 2)$, i.e., the Landau-Zener tunneling probability $P_{LZ}$, as shown in Fig. 3c in the main text, we cannot directly use the asymptotic Landau-Zener formula which is based on sweeping the system across the avoided crossing from negative to positive infinities. In contrast, in our experiment the LZS occurs near the avoided crossings. Therefore, our numerical results are obtained by solving the Bloch equations directly.

\noindent\textbf{Numerical simulation of LZS interference in tripartite qubit-TLS system.} For the tripartite qubit-TLS system discussed below, the qubit is coupled resonantly to two TLSs (TLS1 and TLS2) with different excited state energies $\hbar\omega_{TLS1}$ and $\hbar\omega_{TLS2}$.
The Hamiltonian in the basis of {$|1g_1g_2\rangle$, $|0e_1g_2\rangle$, $|0g_1e_2\rangle$ } is Hamiltonian (\ref{Hamiltonian1}) in the main text. The Bloch equations that govern the evolution of the density operator can be written as: (For simplicity, we
relabel $|1g_1g_2 \rangle$, $|0e_1g_2 \rangle$, $|0g_1e_2 \rangle$ as $|a \rangle$, $|b \rangle$, $|c \rangle$, respectively.)
\begin{equation}\label{2TLS}
\left\{
\begin{array}{rl}
&\dot{\rho}_{aa}=-i\Delta_{1}(\rho_{ba}-\rho_{ab})-i\Delta_{2}(\rho_{ca}-\rho_{ac})-\Gamma_a\rho_{aa}\\
&\dot{\rho}_{bb}= i\Delta_{1}(\rho_{ba}-\rho_{ab})-\Gamma_b\rho_{bb}\\
&\dot{\rho}_{cc}= i\Delta_{2}(\rho_{ca}-\rho_{ac})-\Gamma_c\rho_{cc}\\
&\dot{\rho}_{ab}=-i(\omega_{10}(t)-\omega_{TLS1})\rho_{ab}-i\Delta_{1}(\rho_{bb}-\rho_{aa})\\
&\phantom{\dot{\rho}_{ab}=}-i\Delta_{2}\rho_{cb}-\gamma_{ab}\rho_{ab}\\
&\dot{\rho}_{ac}=-i(\omega_{10}(t)-\omega_{TLS2})\rho_{ac}-i\Delta_{2}(\rho_{cc}-\rho_{aa})\\
&\phantom{\dot{\rho}_{ab}=}-i\Delta_{1}\rho_{bc}-\gamma_{ac}\rho_{ac}\\
&\dot{\rho}_{bc}=-i(\omega_{TLS1}-\omega_{TLS2})\rho_{bc}-i\Delta_{1}\rho_{ac}\\
&\phantom{\dot{\rho}_{ab}=}+i\Delta_{2}\rho_{ba}-\gamma_{bc}\rho_{bc}
\end{array}
\right.
\end{equation}
where the diagonal elements $\rho_{ii}$ are the populations, off-diagonal elements $\rho_{ij}(i\ne j)$ describe coherence, and $\gamma_{ij}=(\Gamma_i+\Gamma_j)/2+\gamma^{(deph)}$ are the rates of decoherence. The remaining three elements' equations are determined by $\rho_{ij}^{\ast}=\rho_{ji}$. The numerically simulated LZS interference pattern is shown in Fig. 3a in the main text, which agrees with experimental result excellently.

\bibliography{nature_comms_LZ}

\begin{addendum}
 \item This work is partially supported by NCET, NSFC (10704034,10725415), 973 Program (2006CB601006), the State Key Program for Basic Research of China (2006CB921801) and NSF Grant No. DMR-0325551.
We acknowledge Northrop Grumman ES in Baltimore MD for technical and foundry support and thank R. Lewis, A. Pesetski, E. Folk, and J. Talvacchio for technical assistance. We thank B. Ruzicka for editing the manuscript.

G.S. and S.H. conceived the experiments; G.S. carried out the measurements with the help of B.M. and analyzed the data with the help of X.W., Y.Y., J.C., P.W. and S.H.; X.W. performed the numerical calculations; G.S., Y.Y. and S.H. wrote the paper.

 \item[Competing Interests] The authors declare that they have no
competing financial interests.
 \item[Correspondence] Correspondence and requests for materials
should be addressed to G.S. (email: gzsun@ku.edu) or Y.Y. (email: yuyang@nju.edu.cn) or S.H. (email: han@ku.edu).
\end{addendum}

\begin{figure}

\caption{\textbf{Qubit circuit and experimental procedure.}
\textbf{a}, Schematic of the qubit circuitry. Josephson
junctions Al/AlOx/Al are denoted by the X symbols. The flux bias, microwave and readout dc-SQUID are inductively coupled to the qubit with inductance $L\approx770 $ pH$ $, capacitance $C\approx240 $ fF$ $ and critical current $I_0\approx1.4 $ $\mu A $.  \textbf{b}, Principle of
the operation and measurement of the phase qubit. The two lowest eigenstates
$|0\rangle $ and $|1\rangle $ form the qubit with transition frequency $\omega
_{10}$ which can be adjusted by changing the flux bias. A microwave pulse is used to manipulate the qubit
state and readout pulse then lower the potential energy barrier to perform a fast single-shot readout. \textbf{c}, Schematic of a two-level state located inside the insulating tunnel barrier of a Josephson junction and its eigenstates in different bases. \textbf{d}, Spectroscopy of the coupled qubit-TLS system with corresponding quantum states labeled. Two avoided crossings centered at $\omega _{TLS1}$ and $\omega _{TLS2}$ are observed. }

\caption{ \textbf{LZS interference in a phase qubit coupled to two
TLSs.} \textbf{a}, The population of $|1\rangle $ measured immediately (a few ns) after the triangular flux
pulse is plotted as a function of the width and amplitude of the triangular flux
bias waveform. The oblique dotted lines are lines of constant characteristic sweeping rate $\nu _{1}$ and $\nu_{2}$ defined in the text. The white circles mark the `hot spots', where the interference fringes generated by $M_{2}$ tend to fade out and the interference fringes generated by $M_{1}$
dominate. \textbf{b} and \textbf{c}, Analytically calculated constructive
interference strips in region I and II, respectively. The horizontal and vertical dotted lines indicate the corresponding locations of interference strips. \textbf{b} and \textbf{c} have the same axis labels as \textbf{a}. \textbf{d}, Schematic of generating LZS
interference with tunable beam splitters in a phase qubit coupled to two TLSs. $M_{1}
$ and $M_{2}$ correspond to the TLSs with smaller and larger avoided crossings in
Fig. 1d, respectively.}

\caption{\textbf{Numerically simulated LZS interference pattern and control of a generalized W state in a phase qubit
coupled to two TLSs.} \textbf{a}, The numerically simulated population of $%
|1\rangle $ after the triangular flux pulse is plotted as a
function of the width and amplitude of the triangular flux bias. The horizontal dotted line indicates the location of $%
\Phi_{LZS}=10 $ m$\Phi _{0}$ and the vertical dotted lines indicate the locations of `hot spots' at $%
\Phi_{LZS}=10 $ m$\Phi _{0}$.
The oblique dotted lines are lines of constant sweeping rate.
The parameters used are determined experimentally: $%
\omega _{01,dc}/2\pi=16.747$ GHz, $|s|=|\frac{\Delta E}{\Delta \Phi}|=0.0404$ GHz/m$%
\Phi _{0}$, $\omega _{TLS1}/2\pi=16.590$ GHz, $\omega _{TLS2}/2\pi=16.510$ GHz, $%
\Delta _{1}/2\pi=10$ MHz, $\Delta
_{2}/2\pi=32$ MHz, $\Gamma _{1g_{1}g_{2}}=(70$ $ns)^{-1}$, $%
\Gamma _{0e_{1}g_{2}}=\Gamma _{0g_{1}e_{2}}=(146$ $ns)^{-1}$, $\gamma
^{(deph)}=(45$ $ns)^{-1}$. \textbf{b}, The upper panel shows the dependence of population
of $|1\rangle$ on $\Phi_{LZS}$ at $T$ = $20$ ns, $40$ ns, $60$ ns, respectively. The lower panel shows
the dependence of population of $|1\rangle$ on $T$
at $\Phi_{LZS}=3.6 $ m$\Phi _{0}, 7.2 $ m$\Phi _{0}, 10.8 $ m$\Phi _{0}$, respectively.
The circles represent the experimental data and the lines from the theory. \textbf{c}, LZ transition probabilities of $M_{1}$ (blue line) and $M_{2}$ (red line) at $%
\Phi_{LZS}=10 $ m$\Phi _{0}$ as a function of pulse width. They are quite different from the asymptotic LZ transition probabilities (blue dotted line and red dotted line). \textbf{d}, The resulting $w$ as a function of $T$ and $\Phi_{LZS}$.}

\end{figure}

\clearpage
\begin{figure}[Fig1.]
\begin{center}
\includegraphics[width=\textwidth]{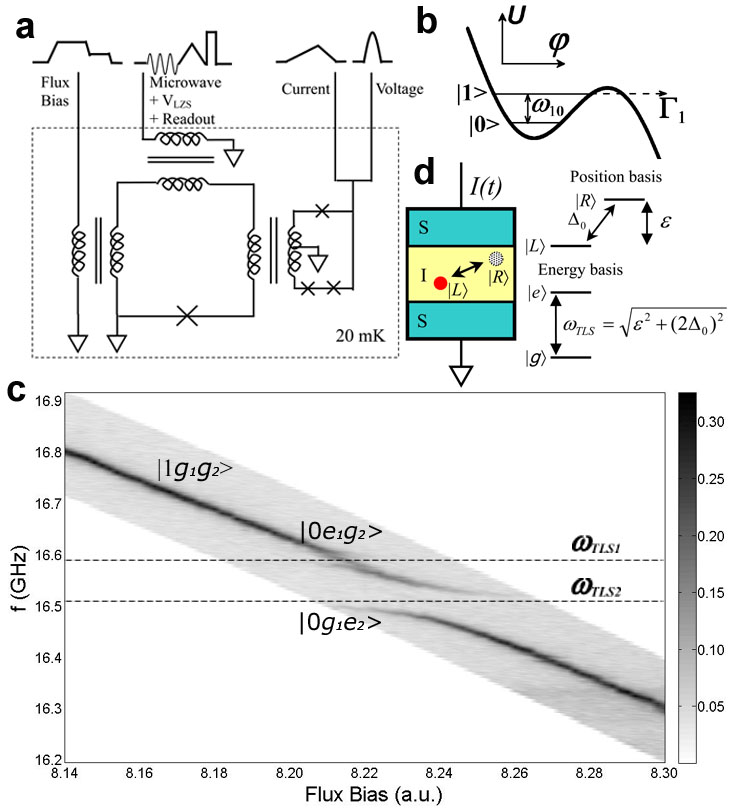}
\end{center}
\end{figure}

\clearpage
\begin{figure}[Fig2.]
\begin{center}
\includegraphics[width=\textwidth]{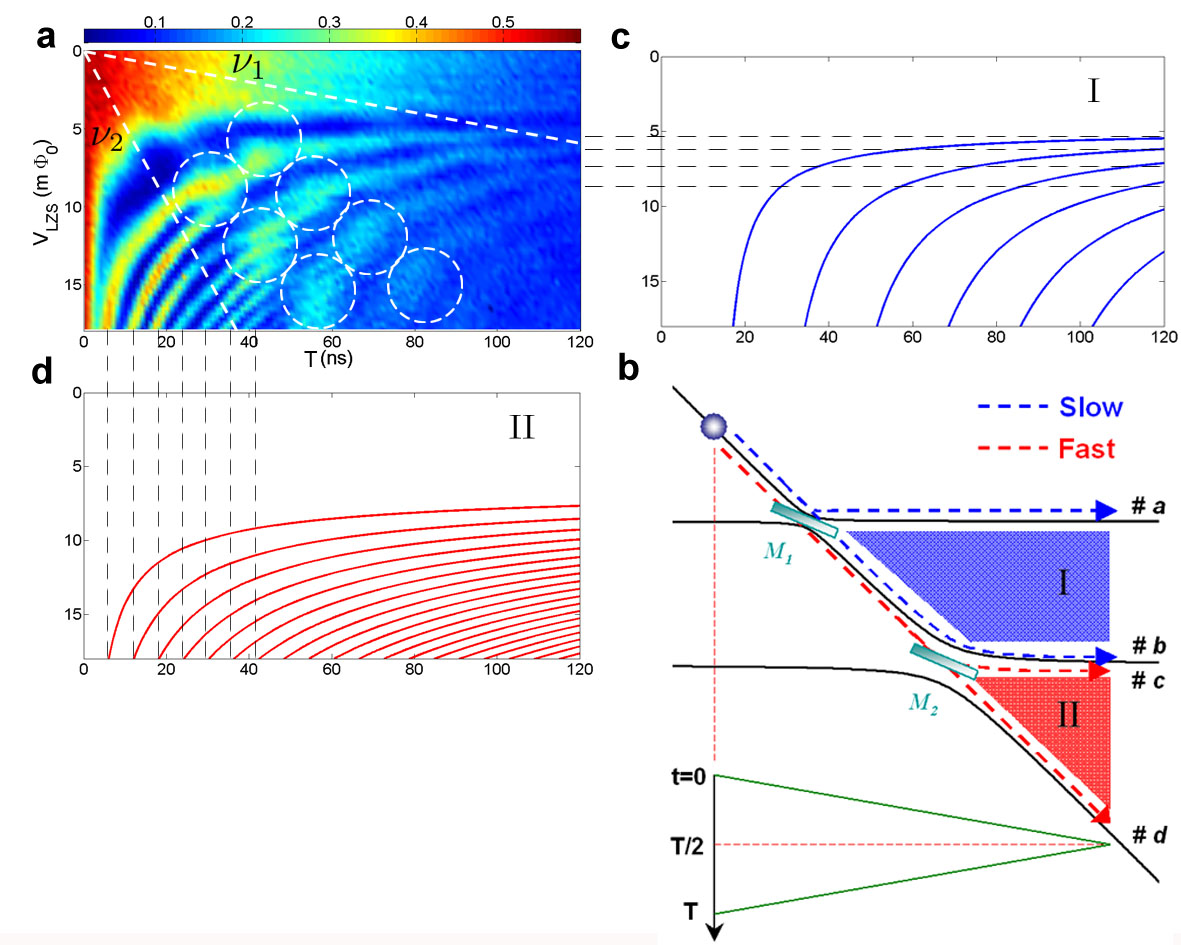}
\end{center}
\end{figure}

\clearpage
\begin{figure}[Fig3.]
\begin{center}
\includegraphics[width=\textwidth]{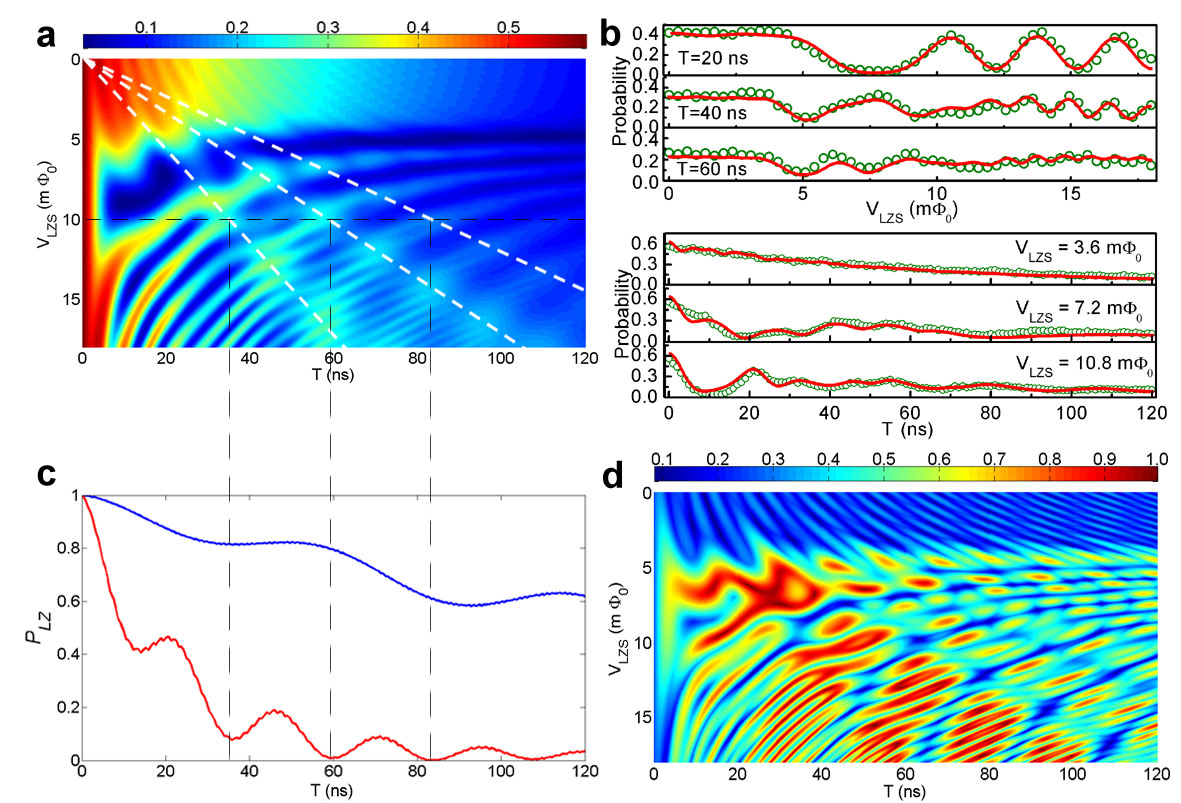}
\end{center}
\end{figure}

\end{document}